%% file: RiskSDPJCompFi.tex
\documentclass{article}
\usepackage{amssymb}
\usepackage{times,mathptmx}
\usepackage[T1]{fontenc}
\usepackage{textcomp}
\usepackage[dvips]{graphicx}
\usepackage{amsmath}

\input defs.tex

\newtheorem{notation}[theorem]{Notation}

\usepackage[agsmcite]{harvard}
\renewcommand{\cite}{\citeasnoun}

\begin{document}

\title{Risk-Management Methods for the Libor Market Model Using Semidefinite
Programming\thanks{This work has been developed under the
direction of Nicole El Karoui and I am very grateful to her. I
also benefited in my work from discussions with Guillaume Amblard,
Marco Avellaneda, Vlad Bally, Stephen Boyd, J\'{e}r\^{o}me Busca,
Rama Cont, Darrell Duffie, Stefano Gallucio, Laurent El Ghaoui,
Jean-Michel Lasry, Marek Musiela, Ezra Nahum, Antoon Pelsser, Yann
Samuelides, Olivier Scaillet, Robert Womersley, seminar
participants at the GdR FIQAM at the Ecole Polytechnique, the May
2002 Workshop on Interest Rate Models organized by Fronti\`{e}res
en Finance in Paris, the June 2002 AFFI conference in Strasbourg
and the Summer School on Modern Convex Optimization at the
C.O.R.E. in U.C.L. Finally, I am also very grateful to
J\'{e}r\^{o}me Lebuchoux, Cyril Godart and everybody inside FIRST
and the S.P.G. at Paribas Capital Markets in London, their advice
and assistance has been key in the development of this work.}}
\author{A. d'Aspremont\thanks{ORFE, Princeton University, Princeton NJ 08544, 
U.S.A. Email: alexandre.daspremont@m4x.org. This work started
while the author was at CMAPX, Ecole Polytechnique, Palaiseau,
France.}} \maketitle

\begin{abstract}
When interest rate dynamics are described by the Libor Market
Model as in \cite{BGM97}, we show how some essential
risk-management results can be obtained from the dual of the
calibration program. In particular, if the objective is to
maximize another swaption's price, we show that the optimal dual
variables describe a hedging portfolio in the sense of
\cite{Avel96}. In the general case, the local sensitivity of the
covariance matrix to all market movement scenarios can be directly
computed from the optimal dual solution. We also show how
semidefinite programming can be used to manage the Gamma exposure
of a portfolio.

\end{abstract}

\section{Introduction}
A recent stream of works on the Libor Market Model have showed how
swap rates can be approximated by a basket of lognormal processes
under an appropriate choice of forward\ measure. This, coupled
with analytic European basket call pricing approximations, allows
to cast the problem of calibrating the Libor Market Model to a set
of European caps and swaptions as a semidefinite program, \ie~a
linear program on the cone of positive semidefinite matrices (see
\cite{Nest94} and \cite{Boyd03}). This work exploits the related
duality theory to provide explicit sensitivity and hedging results
based on the optimal solution to the calibration program.

The lognormal approximation for basket pricing has its origin in
electrical engineering as the addition of noise in decibels (see
for example \cite{schw82}). Its application to basket option
pricing dates back to \cite{Huyn94} or \cite{Musi97}.
\cite{Brac99} tested its empirical validity for swaption pricing
and \cite{Brac00} used it to study Bermudan swaptions. More
recently, \cite{dasp02b}, \cite{Kawa02} and \cite{Kurb02} obtained
additional terms in the expansion and further evidence on the
approximation performance. On the calibration front, \cite{Rebo98}
and \cite{Rebo99} highlight the importance of jointly calibrating
volatilities and correlations. These works, together with
\cite{Long00} also detail some of the most common non-convex
calibration techniques based on parametrizations of the forward
rates covariance factors. The mixed static-dynamic hedging
formulation of the pricing problem has its source in the works by
\cite{ElKa98}, \cite{Avel95} and \cite{Avel96}, while
\cite{Roma00} provide some closed-form results in the multivariate
case.

Here, we show how the dual solution to the calibration program
provides a complete description of the sensitivity to changes in
market condition. In fact, because the algorithms used to solve
the calibration problem jointly solve the problem and its dual,
\textit{the sensitivity of the calibrated covariance matrix is
readily available from the dual solution to the calibration
program}. When the objective in the calibration program is another
swaption's price, the dual solution also describes an approximate
solution to the optimal hedging problem in \cite{Avel96}, which
computes the price of a derivative product as the sum of a static
hedging portfolio and a dynamic strategy hedging the worst-case
residual risk. We also show how semidefinite programming can be
used to efficiently solve the problem of optimally managing the
Gamma exposure of a portfolio using vanilla options, as posed\ by
\cite{Doua95}.

The results we obtain here underline the key advantages of
applying semidefinite programming methods to the calibration
problem: besides their numerical performance, they naturally
provide some central results on sensitivity and risk-management.
They can also eliminate numerical errors in sensitivity
computations that were caused by the inherent instability of
non-convex calibration techniques.

The paper is organized as follows: In the next section, we quickly
recall the calibration program construction for the Libor Market
Model. Section three shows how to compute key sensitivities from
the dual solution to this calibration problem. A fourth section
details how these results can be used to form hedging portfolios.
Finally, in the last section, we present some numerical results.

\section{Model Calibration}
In this section, we begin by briefly recalling the Libor Market
Model construction along the lines of \cite{BGM97} (see also
\cite{Jams97}, \cite{Sand97} and \cite{Milt97}). We then describe
how to form the calibration program.

\subsection{Zero coupon dynamics}
We use the Musiela parametrization of the \cite{Heat92}\ setup.
$r(t,\theta ) $ is the continuously compounded instantaneous
forward rate at time $t$, with duration $\theta $. To avoid any
confusion, Roman letters will be used for maturity dates and Greek
ones for durations. The zero-coupon is here computed as
\begin{equation}
B(t,T)=\exp \left( -\int_{0}^{T-t}r(t,\theta )d\theta \right)  \label{ZCdef}
\end{equation}
\qquad All dynamics are described in a probability space $(\Omega
,\{F_{t};t\geq 0\},\mathbb{P})$ where the filtration $\{F_{t};t\geq 0\}$ is
the $\mathbb{P}$-augmentation of the natural filtration generated by a $d$
dimensional Brownian motion $W=\{W_{t},t\geq 0\}$. The savings account is
defined by:
\begin{equation*}
\beta_{t}=\exp \left( \int_{0}^{t}r(s,0)ds\right)
\end{equation*}
and represents the amount generated at time $t\geq 0$ by continuously
reinvesting 1 euro in the spot rate $r(s,0)$ during the period $0\leq s\leq
t.$ As in \cite{Heat92}, the absence of arbitrage between all zero-coupons
and the savings account states that:
\begin{equation}
\frac{B(t,T)}{\beta_{t}}=B(0,T)\exp \left( -\int_{0}^{t}\sigma
(s,T-s)dW_{s}-\frac{1}{2}\int_{0}^{t}\left| \sigma
(s,T-s)\right|^{2}ds\right)  \label{NoArbCond}
\end{equation}
is a martingale under $\mathbb{P}$ for all $T>0$, where for all $\theta \geq
0$ the zero-coupon bond volatility process $\{\sigma (t,\theta );\theta \geq
0\}$ is $F_{t}$-adapted with values in $\mathbb{R}^{d}$. We assume that the
function $\theta \longmapsto \sigma (t,\theta )$ is absolutely continuous
and the derivative $\tau (t,\theta )=\partial /\partial \theta (\sigma
(t,\theta ))$ is bounded on $\mathbb{R}^{2}\times \Omega .$

\subsection{Libor diffusion process}
All \cite{Heat92} based arbitrage models are fully specified by
their volatility structure and the forward rates curve today. The
specification of the volatility $\sigma (t,\theta )$ in
\cite{BGM97} is based on the assumption that for a given
underlying maturity $\delta$ (for ex. 3 months) the associated
forward Libor process $\{L(t,\theta );t\geq 0\}$ with maturity
$\theta$ defined by:
\begin{equation*}
1+\delta L(t,\theta )=\exp \left( \int_{\theta}^{\theta +\delta
}r(t,\nu )d\nu \right)
\end{equation*}
has a lognormal volatility structure. Using the Ito formula
combined with the dynamics detailed above, we get as in
\cite{BGM97}:
\BEQ
\BA{lll}
dL(s,\theta )& = & \left( \frac{\partial L(s,\theta )}{\partial
\theta }+\frac{\left( 1+\delta L(s,\theta )\right) }{\delta}\sigma^{B}
(s,\theta +\delta)(\sigma^{B}(s,\theta +\delta )-\sigma ^{B}(s,\theta ))\right) ds \\
& & +\frac{1}{\delta }\left( 1+\delta L(s,\theta )\right)
(\sigma^{B}(s,\theta +\delta )-\sigma^{B}(s,\theta ))dW_{s}
\label{LiborVol}
\EA
\EEQ
where $W_s$ is the Brownian motion defined above and the
deterministic volatility function $\gamma
:\mathbb{R}_{+}^{2}\longmapsto \mathbb{R}^{d}$ is bounded and
piecewise continuous. To get the desired lognormal volatility for
Libors we must specify the zero-coupon volatility $\sigma
(t,T_{i})$ as:
\begin{equation}
\sigma (t,T_{i})=\sum_{j=1}^{i-1}\frac{\delta K(t,T_{j})}{1+\delta
K(t,T_{j})}\gamma (t,T_{j}-t)  \label{ZCvol}
\end{equation}
where $\{T_1,\ldots,T_M\}$ is a calendar with period $\delta$ and
the Forward Rate Agreement $K(t,T)=L(t,T-t)$ dynamics are given
by:
\begin{equation*}
dK(s,T)=\gamma(s,T-s)K(s,T)\left[ \sigma
(s,T-s+\delta)ds+dW_{s}\right]
\end{equation*}
Finally, as in \cite{BGM97} we set $\sigma(t,\theta)=0$ for $0\leq
\theta <\delta $.

\subsection{Swaps}
A swap rate is the rate that zeroes the present value of a set of
periodical exchanges of a fixed coupon against a floating coupon
equal to a Libor rate. In a representation that is central in
swaption pricing approximations, we can write swaps as baskets of
forwards (see for ex. \cite{Rebo98}). For example, in the case of
a swap with start date $T_S$ and end date $T_N$:
\begin{equation}
swap(t,T_S,T_{N})=\sum_{i=S}^{N}\omega_{i}(t)K(t,T_{i})
\label{SwapDef}
\end{equation}
where $T_S,T_{N}$ are calendar dates in $\{T_1,\ldots,T_M\}$, and:
\begin{equation}
\omega_{i}(t)=\frac{cvg(T_{i},T_{i+1})B(t,T_{i+1})}{Level(t,T_S,T_{N})}
\label{weightsdef}
\end{equation}
with $cvg(T_{i},T_{i+1})$, the coverage (time interval) between
$T_{i}$ and $T_{i+1}$ and $Level(t,T_S,T_{N})$ the level payment,
\ie~the sum of the discount factors for the fixed calendar of the
swap weighted by their associated coverage:
\begin{equation*}
Level(t,T_S,T_{N})=\sum_{i=S}^{N}cvg(T_{i},T_{i+1})B(t,T_{i+1})
\end{equation*}

\subsection{Swaption price approximation}
As in \cite{Brac00}, \cite{dasp02b} or \cite{Kurb02}, we
approximate the swap dynamics by a one-dimensional lognormal
process, assuming the weights $\omega_{i}(t)$ are constant equal
to $\omega_{i}$:
\begin{equation}
\frac{d~swap(s,T_S,T_{N})}{swap(s,T,T_{N})}=\sum_{i=S}^{N}\hat{\omega}_{i}\gamma
(s,T_{i}-s)dW_{s}^{LVL}  \label{SwapDynamics}
\end{equation}
where
\begin{equation*}
\hat{\omega}_{i}=\omega_{i}\frac{K(t,T_{i})}{swap(t,T_S,T_{N})}
\end{equation*}
is computed from the market data today and $W_{t}^{LVL}$ is a $d$
dimensional Brownian motion under the swap martingale measure
defined in \cite{Jams97}, which takes the level payment as a
num\'{e}raire. We can use the order zero basket pricing
approximation in \cite{Huyn94} and compute the price of a payer
swaption starting with maturity $T,$ written on
$swap(s,T_S,T_{N}),$ with strike $\kappa$ using the \cite{Blac76}
pricing formula:
\begin{equation}
Level(t,T_S,T_{N})\left( swap(t,T_S,T_{N})N(h)-\kappa N\left(
h-\sqrt{V_{T_S}}\right) \right)  \label{swptlogprice}
\end{equation}
where
\begin{equation*}
h=\frac{\left( \ln \left( \frac{swap(t,T_S,T_{N})}{\kappa}\right)
+\frac{1}{2}V_{T_S}\right)}{\sqrt{V_{T_S}}}
\end{equation*}
and $swap(t,T_S,T_{N})$ is the value of the forward swap today
with the variance $V_{T_S}$ given by:
\begin{eqnarray}
V_{T} &=&\int_{t}^{T_S}\left\|
\sum_{i=S}^{N}\hat{\omega}_{i}\gamma (s,T_{i}-s)\right\|^{2}ds
=\int_{t}^{T_S}\left(
\sum_{i,j=S}^{N}\hat{\omega}_{i}\hat{\omega}_{j}\left\langle
\gamma (s,T_{i}-s),\gamma(s,T_{j}-s)\right\rangle \right) ds  \notag \\
&=&\int_{t}^{T_S}\Tr\left( \Omega_{t}\Gamma_{s}\right) ds
\label{varianceformula}
\end{eqnarray}
We note $\symm^{n}$ the set of symmetric matrices of size $n\times
n$. This cumulative variance is a linear form on the forward rates
covariance, with $\Omega_{t}$ and $\Gamma_{s}\in \symm^{N-S+1}$
constructed such that:
\begin{equation*}
\Omega_{t}=\hat{\omega}\hat{\omega}^{T}=\left(
\hat{\omega}_{i}\hat{\omega}_{j}\right)_{i,j\in [S,N]}\succeq 0
~\mbox{ and }~ \Gamma_{s}=\left( \left\langle \gamma
(s,T_{i}-s),\gamma (s,T_{j}-s)\right\rangle \right)_{i,j\in
[S,N]}\succeq 0
\end{equation*}
where $\Gamma_{s}$ is the covariance matrix of the forward rates
(or the Gram matrix of the $\gamma(s,T_{i}-s)$ volatility function
defined above). Here \textit{swaptions are priced as basket
options} with constant coefficients. As detailed in \cite{Brac00}
or \cite{dasp02b}, this simple approximation creates a relative
error on swaption prices of $1-2\%$, which is well within Bid-Ask
spreads. Finally, we remark that caplets can be priced in the same
way as one period swaptions.

\subsection{Semidefinite programming}
In this section we give a brief introduction to semidefinite
programming.
\subsubsection{Complexity}
A standard form linear program can be written:
\[
\BA{ll}
\mbox{minimize}   & c^T x\\
\mbox{subject to} & Ax=b\\
                  & x \succeq 0
\EA
\]
in the variable $x\in\mathbb{R}^n$, where $x \succeq 0$ means here
$x$ componentwise nonnegative. Because their feasible set is the
intersection of an affine subspace with the convex cone of
nonnegative vectors, the objective being linear, these programs
are convex. If the program is feasible, convexity guarantees the
existence of a unique (up to degeneracy or unboundedness) optimal
solution.

The first method used to solve these programs in practice was the
simplex method. This algorithm works well in most cases but is
known to have an exponential worst case complexity. In practice,
this means that convergence of the simplex method cannot be
guaranteed. Since the work of \cite{Nemi79} and \cite{Karm84}
however, we know that these programs can be solved in polynomial
time by interior point methods and most modern solver implement
both techniques.

More importantly for our purposes here, the interior point methods
used to prove polynomial time solvability of linear programs have
been generalized to a larger class of convex problems. One of
these extensions is called semidefinite programming. A standard
form \emph{semidefinite program} is written:
\BEQ
\BA{ll}
\mbox{minimize}   & \Tr(CX)\\
\mbox{subject to} & \Tr(A_iX)=b_i,\quad i=1,\ldots,m\\
                  & X \succeq 0
\EA
\label{standardsdp}
\EEQ
in the variable $X\in\symm^n$, where $X \succeq 0$ means here that
$X$ is positive semidefinite. \cite{Nest94} showed that these
programs can be solved in polynomial time. A number of efficient
solvers are available to solve them, the one used in this work is
called SEDUMI by \cite{Stur99}. In practice, a program with $n=50$
will be solved in less than a second. In what follows, we will
also formulate \emph{semidefinite feasibility problems}:
\[
\BA{ll}
\mbox{find}   &     X\\
\mbox{subject to} & \Tr(A_iX)=b_i,\quad i=1,\ldots,m\\
                  & X \succeq 0
\EA
\]
in the variable $X\in\symm^n$. Their solution set is convex as the
intersection of an affine subspace with the (convex) cone of
positive semidefinite matrices and a particular solution can be
found by choosing an objective matrix $C$ and solving the
corresponding semidefinite program (\ref{standardsdp}). We will
see below that most duality results on linear programming can be
extended to semidefinite programs.

\subsubsection{Semidefinite duality}
We now very briefly summarize the duality theory for semidefinite
programming. We refer again the reader to \cite{Nest94} or
\cite{Boyd03} for a complete analysis. A standard form primal
semidefinite program is written:
\begin{equation}
\BA{ll}
\mbox{minimize} & \Tr(CX) \\
\mbox{s.t.} & \Tr(A_iX)=b_i,\quad i=1,...,m \\
& X \succeq 0
\EA
\end{equation}
in the variable $X\in \symm^{M}$. For $X\succeq 0$, $y\in
\mathbb{R}^{m}$, we form the following Lagrangian:
\begin{align*}
L(X,y)& =\Tr(CX)+\sum_{k=1}^{m}y_{k}\left( b_k-\Tr(A_{k}X)\right) \\
& =\Tr\left( \sum_{k=1}^{m}\left( -y_{k}A_{k}+C\right) X\right)
+\sum_{k=1}^{m}y_{k}b_k
\end{align*}
and because the semidefinite cone is self-dual, we find that
$L(X,y)$ is bounded below in $X\succeq 0$ iff:
\begin{equation*}
C - \sum_{k=1}^{m}y_{k}A_{k} \succeq 0
\end{equation*}
hence the dual semidefinite problem becomes:
\begin{equation}
\BA{ll}
\mbox{maximize} & \sum_{k=1}^{m}y_{k}b_k \\
\mbox{s.t.} &  C-\sum_{k=1}^{m}y_{k}A_{k} \succeq 0
\EA
\end{equation}
When the program is feasible, most solvers produce both primal and
dual solutions to this problem as well as a certificate of
optimality for the solution in the form of the associated duality
gap:
\begin{equation*}
\mu =\Tr\left( X\left(C - \sum_{k=1}^{m}y_{k}A_k\right) \right)
\end{equation*}
which is an upper bound on the absolute error. If on the other
hand the program is infeasible, the dual solution provides a
Farkas type infeasibility certificate (see \cite{Boyd03} for
details). This means that, for reasonably large problems,
semidefinite programming solvers can be used as black boxes.

\subsubsection{Cone programming} The algorithms used to
solve linear and semidefinite programs can be extended a little
further to include \emph{second order cone} constraints. A
\emph{cone program} mixes linear, second order and semidefinite
constraints and is written:
\[
\BA{ll}
\mbox{minimize}   & \Tr(CX)\\
\mbox{subject to} & \|d_j^T\mathop{vec}(X)+e_j\|\leq f_j^T\mathop{vec}(X)+g_j,\quad j=1,\ldots,p\\
                  & \Tr(A_iX)=b_i,\quad i=1,\ldots,m\\
                  & X \succeq 0
\EA
\]
in the variable $X\in\symm^n$, where $\mathop{vec}(X)$ turns $X$
into a vector $x\in\mathbb{R}^{n^2}$ by stacking its columns.
Again, a direct extension of the duality results above is valid
for cone programs and solvers such as SEDUMI by \cite{Stur99} give
either both primal and dual solutions or a certificate of
infeasibility in polynomial time.

\subsection{The calibration program}
Here, we describe the practical implementation of the calibration
program using the swaption pricing approximation detailed\ above.
This is done by discretizing in $s$ the covariance matrix
$\Gamma_{s}$. We suppose that the calibration data set is made of
$m$ swaptions with option maturity $T_{S_{k}} $ written on swaps
of maturity $T_{N_{k}}-T_{S_{k}}$ for $k=1,...,m $, with market
volatility given by $\sigma_{k}$.

\subsubsection{A simple example}
\label{StatExample} In the simple case where the volatility of the
forwards is of the form $\gamma (s,T-s)=\gamma (T-s)$ with $\gamma
$ piecewise constant over intervals of length $\delta $, and we
are given the market price of $k$ swaptions with
$\sigma_{k}^{2}T_{k}\in \mathbb{R}_{+}$ the \cite{Blac76}
cumulative variance of swaption $k$ written on
$swap(t,T_{S_{k}},T_{N_{k}})$, the calibration problem becomes,
using the approximate swaption variance formula in
(\ref{varianceformula}):
\begin{equation}
\BA{ll}
\mbox{find} & X \\
\mbox{s.t.} & \Tr(\Omega_{k}X)=\sigma_{k}^{2}T_{S_{k}}, \quad k=1,...,m \\
& X \succeq 0
\EA
\label{StatCalib}
\end{equation}
which is a semidefinite feasibility problem in the covariance
matrix $X\in \symm^{M}$ and $\Omega_{k}=\sum_{j=1}^{S_{k}}\delta
\varphi_{k,j}$ with $\varphi_{k,j}\in \symm^{M}$ the rank one
matrix with submatrix $\hat{\omega}_{k}\hat{\omega}_{k}^{T}$
starting at element $(j,j)$ and all other blocks equal to zero,
with $M=max_k N_k$. In other words
\[
\left(\varphi_{k,j}\right)_{j+p,j+q}=\hat{\omega}_{k,p}\hat{\omega}_{k,q},\quad
p,q=1,\ldots,N_k-S_k+1
\]
with all other elements equal to zero. See \cite{Brac99} or
\cite{dasp02b} for further details.

\subsubsection{The general case}
Here we show that for general volatilities $\gamma(s,T-s)$, the
format of the calibration problem remains similar to that of the
simple example above, except that $X$ becomes block-diagonal. In
the general non-stationary case where $\gamma$ is of the form
$\gamma(s,T-s)$ and piecewise constant on intervals of size
$\delta$, the expression of the market cumulative variance
becomes, according to formula (\ref{varianceformula}):
\begin{equation*}
\sigma_{k}^{2}T_{S_{k}}=\sum_{i=1}^{S_{k}}\delta \Tr\left(
\Omega_{k,i}X_{i}\right) =\int_{t}^{T_S}\left(
\sum_{i,j=S}^{N}\hat{\omega}_{i}\hat{\omega}_{j}\left\langle
\gamma (s,T_{i}-s),\gamma(s,T_{j}-s)\right\rangle \right) ds
\end{equation*}
where $\Omega_{k,i}\in \symm^{M-i}$ is\ a block-matrix with
submatrix $\hat{\omega}_{k}\hat{\omega}_{k}^{T}$ starting at
element $(S_{k}-i,S_{k}-i)$ and all other blocks equal to zero.
Here $X_{i}$ is the Gram matrix of the vectors $\gamma
(T_{i},T_{j}-T_{i})$. Calibrating the model to the swaptions
$k=(1,...,m)$ can then be written as the following semidefinite
feasibility problem:
\begin{equation*}
\BA{ll}
\mbox{find} & X_{i} \\
\mbox{s.t.} & \sum_{i=1}^{S_{k}}\delta \Tr\left(
\Omega_{k,i}X_{i}\right)=\sigma_{k}^{2}T_{S_k}, \quad k=1,...,m \\
& X_{i}\succeq 0,\quad i=1,...,M
\EA
\end{equation*}
with variables $X_{i}\in \symm^{M-i}$. We can write this general
problem in the same format used in the simple stationary case. Let
$X$ be the block matrix
\begin{equation*}
X=\left[
\begin{array}{cccc}
X_{1} & 0 & . & 0 \\
0 & . & . & . \\
. & . & . & 0 \\
0 & . & 0 & X_{M}
\end{array}
\right]
\end{equation*}
the calibration program can be written as in (\ref{StatCalib}):
\begin{equation}
\BA{ll}
\mbox{find} & X \\
\mbox{s.t.} & \Tr(\bar{\Omega}_{k}X)=\sigma_{k}^{2}T_{S_k},\quad k=1,...,m \\
& X\succeq 0,~X\mbox{ block-diagonal}
\EA
\end{equation}
except that $\bar{\Omega}_{k}$ and $X$ are here ''block-diagonal''
matrices. We can also replace the equality constraints with
Bid-Ask spreads. In the simple case detailed in (\ref{StatCalib}),
the new calibration problem is then written as the following
semidefinite feasibility problem:
\begin{equation*}
\BA{ll}
\mbox{find} & X \\
\mbox{s.t.} & \sigma_{Bid,k}^{2}T_{S_{k}}\leq \Tr(\Omega_{k}X)\leq \sigma_{Ask,k}^{2}T_{S_{k}},\quad k=1,...,m \\
& X\succeq 0
\EA
\end{equation*}
in the variable $X\in \symm^{M}$, with parameters $\Omega_{k},$
$\sigma_{Bid,k}^{2},$ $\sigma_{Ask,k}^{2},$ $T_{S_{k}}$. Again, we
can rewrite this program as a semidefinite feasibility problem:
\begin{equation*}
\BA{ll}
\mbox{find} & X \\
\mbox{s.t.} & \Tr\left( \left[
\begin{array}{ccc}
\Omega_{k} & 0 & 0 \\
0 & I & 0 \\
0 & 0 & 0
\end{array}
\right] \left[
\begin{array}{ccc}
X & 0 & 0 \\
0 & U_{1} & 0 \\
0 & 0 & U_{2}
\end{array}
\right] \right) =\sigma_{Ask,k}^{2}T_{S_{k}} \\
&  \\
& \Tr\left( \left[
\begin{array}{ccc}
\Omega_{k} & 0 & 0 \\
0 & 0 & 0 \\
0 & 0 & -I
\end{array}
\right] \left[
\begin{array}{ccc}
X & 0 & 0 \\
0 & U_{1} & 0 \\
0 & 0 & U_{2}
\end{array}
\right] \right) =\sigma_{Bid,k}^{2}T_{S_{k}},\quad k=1,...,m \\
&  \\
& X,U_{1},U_{2}\succeq 0
\EA
\end{equation*}
in the variables $X,U_1,U_2\in\symm^M$, which can be summarized as
\begin{equation}
\BA{ll}
\mbox{find} & \tilde{X} \\
\mbox{s.t.} & \Tr(\tilde{\Omega}_{Ask,k}\tilde{X})=\sigma_{Ask,k}^{2}T_{S_{k}} \\
&
\Tr(\tilde{\Omega}_{Bid,k}\tilde{X})=\sigma_{Bid,k}^{2}T_{S_{k}}\quad k=1,...,m \\
& \tilde{X}\succeq 0,~\tilde{X} \mbox{ block-diagonal}
\EA
\end{equation}
with $\tilde{X},\tilde{\Omega}_{k}\in \symm^{3M}.$ Because of
these transformations and to simplify the analysis, we will always
discuss the stationary case with equality constraints
(\ref{StatCalib}) in the following section, knowing that all
results can be directly extended to the general case using the
block-diagonal formulation detailed above.

\section{Sensitivity analysis}
In this section, we show how the dual optimal solution can be
exploited for computing solution sensitivities with minimal
numerical cost.

\subsection{Computing sensitivities}
Let us suppose that we have solved both the primal and the dual
calibration problems above with market constraints
$\sigma_{k}^{2}T_{S_{k}}$ and let $X^{\mathrm{opt}}$ and
$y^{\mathrm{opt}}$ be the optimal solutions. Suppose also that the
market swaption price constraints are modified by a small amount
$u\in \mathbb{R}^{m}$. The new calibration problem becomes:
\begin{equation}
\label{calib-sensitivity}
\BA{ll}
\mbox{minimize} & \Tr(CX) \\
\mbox{s.t.} & \Tr(\Omega_{k}X)=\sigma_{k}^{2}T_{S_{k}}+u_{k}\quad k=1,...,m \\
& X\succeq 0
\EA
\end{equation}
in the variable $X\in \symm^{M}$ with parameters $\Omega_{k},C\in
\symm^{M}$ and $\sigma_{k}^{2}T_{S_{k}}+u_k\in \mathbb{R}_{+}$.
Here $C$ is, for example, an historical estimate of the covariance
matrix. If we note $p^{\mathrm{opt}}(u)$ the optimal solution to
the revised problem, we get (at least formally for now) the
sensitivity of the solution to a change in market condition as:
\begin{equation}
\frac{\partial p^{\mathrm{opt}}(0)}{\partial
u_{k}}=-y_{k}^{\mathrm{opt}}
\end{equation}
where $y^{\mathrm{opt}}$ is the optimal solution to the dual
problem (see \cite{Boyd03} for details). As we will see in this
section and the next one, this has various interpretations
depending on the objective function. Here, we want to study the
variation in the solution matrix $X^{\mathrm{opt}}$, given a small
change $u$ in the market conditions.

We start with the following definition.
\begin{notation}
Let us suppose that we have solved the general calibration problem
in (\ref{calib-sensitivity}), we call $X^{\mathrm{opt}}$ and
$y^{\mathrm{opt}}$ the primal and dual solutions to the above
problem with $u=0$. We note
\begin{equation*}
Z^{\mathrm{opt}}=\left(
C-\sum_{k=1}^{m}y_{k}^{\mathrm{opt}}\Omega_{k}\right)
\end{equation*}
the dual solution matrix. As in \cite{Aliz98}, we also define the symmetric
Kronecker product as:
\begin{equation*}
\left( P\circledast Q\right) K:=\frac{1}{2}\left( PKQ^{T}+QKP^{T}\right)
\end{equation*}
We note $A$ and $A^{\ast}$, the linear operators defined by:
\begin{equation*}
\begin{array}{l}
A:\symm^{M}\longrightarrow \mathbb{R}^{m} \\
X\longmapsto AX:=\left( \Tr\left(\Omega_{k}X\right)
\right)_{k=1,...,m}
\end{array}
\mbox{ and its dual }
\begin{array}{l}
A^{\ast}:\mathbb{R}^{m}\longrightarrow \symm^{M} \\
y\longmapsto A^{\ast}y:=\sum_{k=1}^{m}y_{k}\Omega_{k}
\end{array}
\end{equation*}
\end{notation}

We now follow \cite{Todd99} to compute the impact $\Delta X$ on
the solution of a small change in the market price data $\left(
u_{k}\right)_{k=1,..,m}$, \ie~given $u$ small enough we compute
the next Newton step $\Delta X$. Each solver implements one
particular search direction to compute this step and we define a
matrix $M$, with $M=I$ in the case of the A.H.O. search direction
based on the work by \cite{Aliz98}, see \cite{Todd99} for other
examples. We also define the linear operators:
\begin{equation*}
E=Z^{\mathrm{opt}}\circledast M\mbox{ and
}F=MX^{\mathrm{opt}}\circledast I
\end{equation*}
and their adjoints
\begin{equation*}
E^{\ast}=Z^{\mathrm{opt}}\circledast M\mbox{ and
}X^{\mathrm{opt}}M\circledast I
\end{equation*}
We remark that if $A,B\in \symm^{M}$commute, with eigenvalues
$\alpha ,\beta \in \mathbb{R}^{M}$ and common eigenvectors $v_{i}$
for $i=1,...,M$, then $A\circledast B$ has eigenvalues
$(\alpha_{i}\beta_{j}+\alpha_{j}\beta_{i})$ for $i,j=1,...,M$ and
eigenvectors $v_{i}v_{i}^{T}$ if $i=j$ and
$(v_{i}v_{j}^{T}+v_{j}v_{i}^{T})$ if $i\neq j$ for $i,j=1,...,M$.
Provided the strict feasibility and nonsingularity conditions in
\S 3 of \cite{Todd99} hold, we can compute the Newton step $\Delta
X$ as:
\begin{equation}
\Delta X=E^{-1}FA^{\ast}\left[ \left( AE^{-1}FA^{\ast
}\right)^{-1}u\right] \label{NewtonStep}
\end{equation}
and this will lead to a feasible point $X^{\mathrm{opt}}+\Delta
X\succeq 0$ iff the market variation movement $u$ is such that:
\begin{equation}
\left\| \left( X^{\mathrm{opt}}\right)^{-\frac{1}{2}}\left(
E^{-1}FA^{\ast }\left[ \left( AE^{-1}FA^{\ast}\right)^{-1}u\right]
\right) \left(
X^{\mathrm{opt}}\right)^{-\frac{1}{2}}\right\|_{2}\leq 1
\label{PositivityRegion}
\end{equation}
where $\|\cdot\|_2$ is the $l_2$ norm. The intuition behind this
formula is that semidefinite programming solvers are based on the
Newton method and condition (\ref{PositivityRegion}) ensures that
the solution $X^{\mathrm{opt}}$ remains in the region of quadratic
convergence of the Newton algorithm.\ This means that only one
Newton step is required to produce the new optimal solution
$X^{\mathrm{opt}}+\Delta X$ and (\ref{NewtonStep}) simply computes
this step. The matrix in (\ref{NewtonStep}) produces a
\textit{direct method for updating }$X$ which we can now use to
compute price sensitivities for any given portfolio.

This illustrates how a semidefinite programming based calibration
allows to test various \textit{realistic scenarios} at a minimum
numerical cost and improves on the classical non-convex methods
that either had to ''bump the market data and recalibrate'' the
model for every scenario with the risk of jumping from one local
optimum to the next, or simulate unrealistic market movements by
directly adjusting the covariance matrix. One key question is
stability: the calibration program in (\ref{StatCalib}) has a
unique solution, but as usual this optimum can be very unstable
and the matrix in (\ref{NewtonStep}) badly conditioned. In the
spirit of the work by \cite{Cont01} on volatility surfaces, we
look in the next section for a way to solve this conditioning
issue and stabilize the calibration result.

\subsection{Robustness}
The previous sections were focused on how to compute the impact of
a change in market conditions. Here we will focus on how to
anticipate those variations and make the calibrated matrix robust
to a given set of scenarios. Depending on the way the
perturbations are modelled, this problem can remain convex and be
solved very efficiently. Let us suppose here that we want to solve
the calibration problem on a set of market Bid-Ask spreads data. A
direct way to address the conditioning issues detailed in the last
section is to use a Tikhonov stabilization of the calibration
program as in \cite{Cont01} and solve the following cone program:
\begin{equation}
\label{tik-stabilization}
\BA{ll}
\mbox{minimize} & \|X\|_2 \\
\mbox{s.t.} & \sigma_{Bid,k}^{2}T_{S_{k}}\leq \Tr(\Omega_{k}X)\leq \sigma_{Ask,k}^{2}T_{S_{k}},\quad k=1,...,m \\
& X\succeq 0
\EA
\end{equation}
in the variable $X\in \symm^{M}$ with parameters $\Omega_{k},C\in
\symm^{M}$ and $\sigma_{Bid,k}^{2}T_{S_{k}}$,
$\sigma_{Ask,k}^{2}T_{S_{k}}\in \mathbb{R}_{+}$. In the absence of
any information on the uncertainty in the market data, we can
simply maximize the distance between the solution and the market
bounds to ensure that it remains valid in the event of a small
change in the market variance input. As the robustness objective
is equivalent to a distance maximization between the solution and
the constraints (or Chebyshev centering), the input of assumptions
on the movement structure is equivalent to a choice of norm.
Without any particular structural information on the volatility
market dynamics, we can use the $l_{\infty}$ norm and the
calibration problem becomes:
\begin{equation*}
\BA{ll}
\mbox{maximize} & t \\
\mbox{s.t.} & \sigma_{Bid,k}^{2}T_{S_{k}}+t\leq
\Tr(\Omega_{k}X)\leq
\sigma_{Ask,k}^{2}T_{S_{k}}-t,\quad k=1,...,m \\
& X\succeq 0
\EA
\end{equation*}
in the variables $t\in\mathbb{R}$ and $X\in\symm^M$. Using the
$l_{1}$ norm instead, this becomes:
\begin{equation*}
\BA{ll}
\mbox{maximize} & \sum_{i=1}^{m}t_{k} \\
\mbox{s.t.} & \sigma_{Bid,k}^{2}T_{S_{k}}+t_{k}\leq
\Tr(\Omega_{k}X)\leq
\sigma_{Ask,k}^{2}T_{S_{k}}-t_{k},\quad k=1,...,m \\
& X\succeq 0
\EA
\end{equation*}
in the variables $t_1,\ldots,t_m\in\mathbb{R}$ and $X\in\symm^M$.
The problems above optimally center the solution within the
Bid-Ask spreads, which makes it robust to a change in market
conditions given no particular information on the nature of that
change. In the same vein, \cite{Hsdp98} also show how to design a
program that is robust to a change in the matrices $\Omega_{k}$.
However, because the matrices $\Omega_{k}$ are computed from
ratios of zero-coupon bonds, their variance is negligible in
practice compared to that of $\sigma_{k}^{2}.$

Suppose now$\ $that $V$ is a statistical estimate of the daily
covariance of $\sigma_{k}^{2}T_{S_{k}}$ (the mid-market
volatilities in this case) and let us assume that these
volatilities have a Gaussian distribution. We adapt the method
used by \cite{Lobo98} for robust L.P. We suppose that the matrix
$V$ has full rank. We can then calibrate the model to this
information:
\begin{equation*}
\BA{ll}
\mbox{maximize} & \Tr(CX) \\
\mbox{s.t.} & \Tr\left( \Omega_{k}X\right) -\sigma_{k}^{2}T_{S_{k}}=v_{k}\mbox{ \ for}k=1,...,m \\
& \left\| V^{-\frac{1}{2}}v\right\|_{\infty}\leq \Phi^{-1}(\mu ) \\
& X\succeq 0
\EA
\end{equation*}
where $\left\| \cdot \right\|_{\infty}$ is the $l_{\infty}$ norm
and $\Phi (x)$ is given by
\begin{equation*}
\Phi (x)=1-\frac{1}{\sqrt{2\pi}}\int_{-x}^{x}\exp (-u^{2}/2)du
\end{equation*}
There is no guarantee that this program is feasible and we can solve instead
for the best confidence level by forming the following program:
\begin{equation*}
\begin{array}{ll}
\mbox{minimize} & t \\
\mbox{s.t.} & \Tr\left( \Omega_{k}X\right) -\sigma_{k}^{2}T_{S_{k}}=v_{k}\mbox{ \ for}k=1,...,m \\
& \left\| V^{-\frac{1}{2}}v\right\|_{\infty}\leq t \\
& X\succeq 0
\end{array}
\end{equation*}
The optimal confidence level is then $\eta =\Phi (t)$ and
``centers'' the calibrated matrix with respect to the uncertainty
in $\sigma_{k}^{2}T_{S_{k}}$. This is a symmetric cone program and
can be solved efficiently.

\section{Hedging}
In this section, we show how semidefinite programming calibration
techniques can be used to build a superreplicating portfolio,
approximating the upper and lower hedging prices defined in
\cite{Karou95}. An efficient technique for computing those price
bounds with general non-convex payoffs on a single asset with
univariate dynamics was introduced in \cite{Avel95} and recent
work on this topic by \cite{Roma00} provided closed-form solutions
for the prices of exchange options and options on the geometrical
mean of two assets.

\subsection{Approximate solution}
\label{sec:hedging-approx} Here, using the approximation in
(\ref{swptlogprice}), we first compute arbitrage bounds on the
price of a basket by adapting the method used by \cite{Avel95} in
the one-dimensional case. We then provide approximate solutions
for these arbitrage bounds on swaptions and show how one can use
the dual solution to build an optimal hedging portfolio in the
sense of \cite{Avel96}, using derivative securities taken from the
calibration set.

As in \cite{Avel96}, the price here is derived from a mixed static-dynamic
representation:
\begin{equation}
\mbox{\textbf{Price} = Min}\left\{ \mbox{Value of \textbf{static
hedge}}+\mbox{Max}\left( \mbox{PV of \textbf{residual
liability}}\right) \right\} \label{PricingEquation}
\end{equation}
where the static hedge is a portfolio composed of the calibration
assets and the maximum residual liability is computed as in
\cite{ElKa98} or \cite{Avel95}. This hedging representation
translates into the pricing problem the market habit of
calibrating a model on the instrument set that will be used in
hedging. The static portfolio uses these instrument to reduce the
payoff risk, while the dynamic hedging part hedges the remaining
risk in conservative manner.

Furthermore, because of the sub-additivity of the above program
with respect to payoffs, we expect this diversification of the
volatility risk to bring down the total cost of hedging. Let
$K(t)=\left( K(t,T_{i})\right)_{i=1,...,M}$ and suppose we have a
set of market prices $C_{k}$ for $k=1,\ldots ,m$, of swaptions
with corresponding market volatilities $\sigma_{k}$, coefficient
matrices $\Omega_{k}\in \symm^{M}$ and payoffs $h_k(K(t))$. As in
\cite{Avel96} we can write the price (\ref{PricingEquation}) of an
an additional swaption with payoff $h_0(K(t))$:
\begin{equation}
\inf_{\lambda \in \mathbb{R}^{m}}\left\{
\sum_{k=1}^{m}\lambda_{k}C_{k}+\left( \sup_{P}E^{P}\left[ \beta
(T)^{-1}h_{0}(K(T))-\sum_{k=1}^{m}\lambda_{k}\beta
(T)^{-1}h_{k}(K(T))\right] \right) \right\}
\end{equation}
where $P$ varies within the set of equivalent martingale measure and $\beta
(T)$ is the value of the savings account in $T$. We can rewrite the above
problem as:
\begin{equation*}
\inf_{\lambda}\left\{ \sup_{P}\left( E^{P}\left[ \beta
(T)^{-1}h_{0}(K(T))\right] -\sum_{i=1}^{m}\lambda_{k}\left(
E^{P}\left[ \beta (T)^{-1}h_{k}(K(T))\right] -C_{k}\right) \right)
\right\}
\end{equation*}
where we recognize the optimum hedging portfolio problem as the dual of the
maximum price problem above:
\begin{equation*}
\BA{ll}
\mbox{maximize} & E^{P}\left[ \beta (T)^{-1}h_{0}(K(T))\right]  \\
\mbox{s.t.} & E^{P}\left[ \beta (T)^{-1}h_{k}(K(T))\right]
=C_{k},\quad k=1,...,m
\EA
\end{equation*}
Using (\ref{swptlogprice}), we get an approximate solution by solving the
following problem:
\begin{equation*}
\BA{ll}
\mbox{maximize} & BS_{0}(\Tr(\Omega_{0}X)) \\
\mbox{s.t.} & BS_{k}\left( \Tr(\Omega_{k}X)\right) =C_{k},\quad k=1,...,m \\
& X\succeq 0 \EA
\end{equation*}
and its dual:
\begin{equation*}
\inf_{\lambda}\left\{ \sup_{X\succeq 0}\left(
BS(\Tr(\Omega_{0}X))-\sum_{k=1}^{m}\lambda_{k}\left( BS\left(
\Tr(\Omega_{k}X)\right) -C_{k}\right) \right) \right\}
\end{equation*}
The primal problem, after we write it in terms of variance, becomes the
following semidefinite program:
\begin{equation*}
\BA{ll}
\mbox{maximize} & \sigma_{\max}^{2}T=\Tr(\Omega_{0}X) \\
\mbox{s.t.} & \Tr(\Omega_{k}X)=\sigma_{k}^{2}T_{k},\quad k=1,...,m
\\
& X\succeq 0
\EA
\end{equation*}
Again, we note $y^{\mathrm{opt}}\in \mathbb{R}^{m}$ the solution
to the dual of this last problem:
\begin{equation*}
\BA{ll}
\mbox{minimize} & \sum_{k=1}^{m}y_{k}\sigma_{k}^{2}T_{k} \\
\mbox{s.t.} & 0\preceq \sum_{k=1}^{m}y_{k}\Omega_{k}-\Omega_{0}
\EA
\end{equation*}
The KKT optimality\ conditions on the primal-dual semidefinite
program pair above (see \cite{Boyd03} for example) can be written:
\begin{equation*}
\left\{
\begin{array}{l}
0\preceq \sum_{k=1}^{m}y_{k}\Omega_{k}-\Omega_{0} \\
0=\sum_{k=1}^{m}y_{k}\Omega_{k}X-\Omega_{0}X \\
\Tr(\Omega_{k}X)=\sigma_{k}^{2}T_{k},\quad k=1,...,m \\
0\preceq X
\end{array}
\right.
\end{equation*}
and we can compare those to the KKT conditions for the price maximization
problem:
\begin{equation*}
\left\{
\begin{array}{l}
Z=\frac{\partial BS_{0}\left( \Tr(\Omega_{0}X)\right)}{\partial
v}\Omega_{0}+\sum_{k=1}^{m}\lambda_{k}\frac{\partial BS_{k}\left(
\Tr(\Omega_{k}X)\right)}{\partial v}\Omega_{k} \\
XZ=0 \\
BS_{k}\left( \Tr(\Omega_{k}X)\right) =C_{i},\quad k=1,...,m \\
0\preceq X,Z
\end{array}
\right.
\end{equation*}
with dual variables $\lambda \in \mathbb{R}^{m}$ and $Z\in
\symm_{n}$. An optimal dual solution $\lambda_{k}^{\mathrm{opt}}$
for the price maximization problem can then be constructed from
$y^{\mathrm{opt}}$, the optimal dual solution of the semidefinite
program on the variance, as:
\begin{equation*}
\lambda_{k}^{\mathrm{opt}}=-y_{k}^{\mathrm{opt}}\frac{\partial
BS_{0}\left( \Tr(\Omega_{0}X)\right) /\partial v}{\partial
BS_{k}\left( \Tr(\Omega_{k}X)\right) /\partial v}
\end{equation*}
which gives the amount of basket $k$ in the optimal static hedging
portfolio defined by (\ref{PricingEquation}) .

\subsection{The exact problem}

The bounds found in the section above are only approximate
solutions to the superreplicating problem. Although the relative
error in the swaption price approximation is known to be about
1-2\%, it is interesting to notice that while being somewhat
intractable, the exact problem shares the same optimization
structure as the approximate one. Let us recall the results in
\cite{Roma00}. If, as above, we note $C(K(t),t)$ the
superreplicating price of a basket option, then $C(K(t),t)$ is the
solution to a multidimensional Black-Scholes-Barenblatt (BSB)
equation. We can create a superreplicating strategy by dynamically
trading in a portfolio composed of $\Delta _{t}^{i}=\frac{\partial
C}{\partial x_{i}}(t,K(t,T_{i}))$ in each asset. The BSB equation
in \cite{Roma00} can be rewritten in a format that is similar to
that of the approximate problem above, to become:
\begin{equation*}
\left\{
\begin{array}{l}
\frac{\partial C(x,t)}{\partial t}+\frac{1}{2}\max_{\Gamma \in
\Lambda}\Tr\left( diag(x)\frac{\partial^{2}C(x,t)}{\partial
x^{2}}diag(x)\Gamma
\right) =0 \\
\\
C(x,T)=\left( \sum_{i=1}^{n}\omega_{i}x_{i}-k\right)^{+}
\end{array}
\right.
\end{equation*}
where $diag(x)$ is the diagonal matrix formed with the components
of $x$ and $\Gamma =\gamma \gamma^{T}$ is the model covariance
matrix. If the set $\Lambda $ is given by the intersection of the
semidefinite cone (the covariance matrix has to be positive
semidefinite) with a polyhedron (for example approximate price
constraints, sign constraints or bounds on the matrix
coefficients, ...), then the embedded optimization problem in
(\cite{Roma00}) becomes a semidefinite program:
\begin{equation*}
\max_{\Gamma \in \Lambda}\Tr\left( \Gamma
diag(x)\frac{\partial^{2}C(x,t)}{\partial x^{2}}diag(x)\right)
\end{equation*}
on the feasible set $\Lambda $. We recover the same optimization
problem as in the approximate solution found in the section above,
the only difference being here that the solution to the general
problem might not be equal to a Black-Scholes price. This gives a
simple interpretation of the embedded optimization problem in the
BSB equation developed in \cite{Roma00}.

\subsection{Optimal Gamma Hedging}
For simplicity here, we work in a pure equity framework and, along
the lines of \cite{Doua95}, we study the problem of optimally
adjusting the Gamma of a portfolio using only options on single
assets. This problem is essentially motivated by a difference in
liquidity between the vanilla and basket option markets, which
makes it impractical to use some baskets to adjust the Gamma of a
portfolio. Suppose we have an initial portfolio with a Gamma
sensitivity matrix given by $\Gamma$ in a market with underlying
assets $x_{i}$ for $i=1,\ldots n$. We want to hedge (imperfectly)
this position with $y_{i} $ vanilla options on each single asset
$x_{i}$ with Gamma given by $\gamma_{i}$. We assume that the
portfolio is maintained delta-neutral, hence a small perturbation
of the stock price will induce a change in the portfolio price
given by:
\begin{equation*}
\Delta P(x+\Delta x)=P(x)+\frac{1}{2}\Delta x^{T}\Gamma (y)\Delta
x
\end{equation*}
where $\Gamma (y)=\Gamma +diag(\gamma )y,$ with $diag(\gamma )$
the diagonal matrix with components $\gamma_{i}$. As in
\cite{Doua95}, our objective is to minimize in $y$ the maximum
possible perturbation given by:
\begin{equation*}
\max_{\Delta x\in \mathcal{E}}\left| \Delta x^{T}\Gamma (y)\Delta
x\right|
\end{equation*}
where $\mathcal{E}$ is the ellipsoid defined by
\begin{equation*}
\mathcal{E}=\left\{ u\in \reals^{n}|u^{T}\Sigma u=1\right\}
\end{equation*}
with $\Sigma_{ij} =cov(x_{i},x_{j})$ for $i,j=1,\ldots,n$, the
covariance matrix of the underlying assets. This amounts to
minimizing the maximum eigenvalue of the matrix $\Sigma \Gamma
(y)$ and can be solved by the following semidefinite program:
\begin{equation*}
\begin{array}{ll}
\mbox{minimize} & t \\
\mbox{subject to} & -tI\preceq \Sigma \Gamma +\Sigma
diag(\gamma)diag(y)\preceq tI
\end{array}
\end{equation*}
in the variables $t\in\mathbb{R}$ and $y\in\mathbb{R}^n$.

\begin{figure}[p]
\begin{center}
\includegraphics[width=1.0\textwidth]{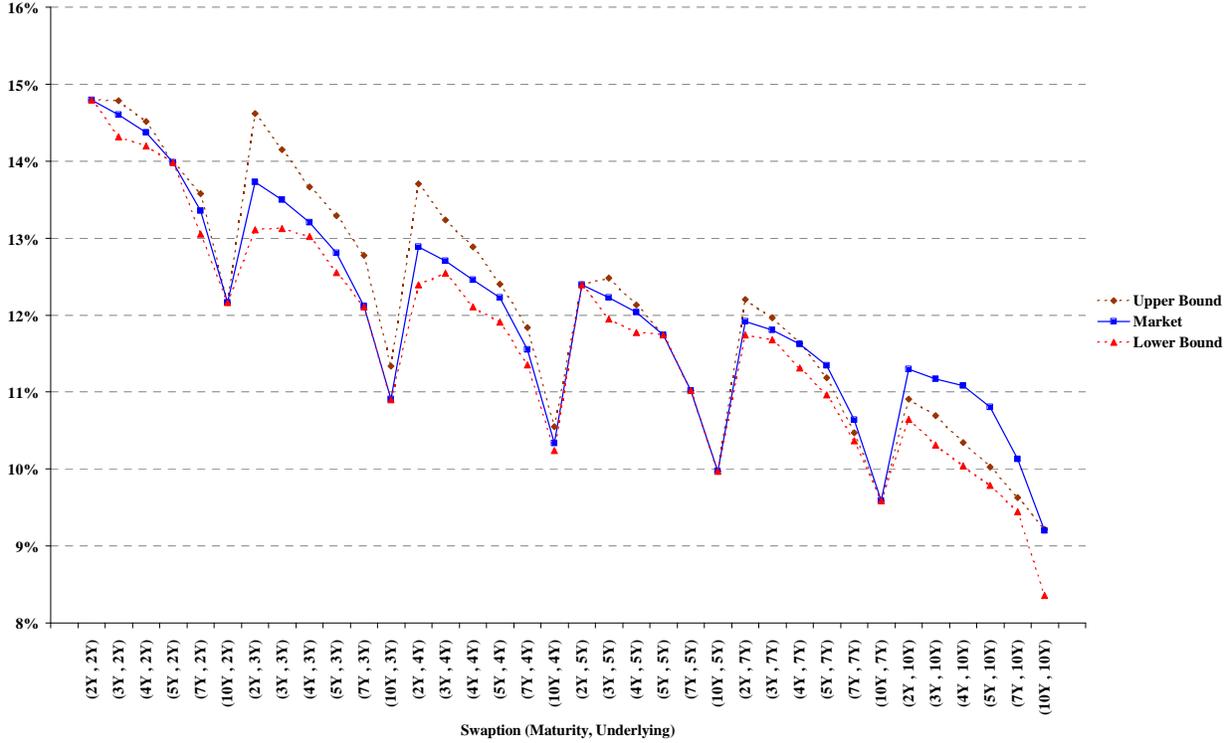}
\end{center}
\caption{Calibration result and price bounds on a ''Sydney opera house'' set
of swaptions.}
\label{figsidney}
\end{figure}
\begin{table}[p]
\begin{center}
\begin{tabular}{lllllllllll}
\textbf{Caplet Vols (\%, 1Y to 10Y)} & 14.3 & 15.6 & 15.4 & 15.1 & 14.8 &
14.5 & 14.2 & 14.0 & 13.9 & 13.3 \\
\textbf{Caplet Vols (\%, 11Y to 20Y)} & 13.0 & 12.7 & 12.4 & 12.2 & 12.0 &
11.9 & 11.8 & 11.8 & 11.7 & 12.0
\end{tabular}
\end{center}
\caption{Caplet volatilities.}
\label{TabCaplets}
\end{table}
\begin{table}[p]
\begin{center}
\begin{tabular}{ccccccccc}
\textbf{Swaption} & \textbf{Vol (\%)} & $\hat{\omega}_{i}$ &  &  &  &  &  &
\\ \hline
2Y into 5Y & 12.4 & 0.22 & 0.20 & 0.20 & 0.19 & 0.18 &  &  \\
5Y into 5Y & 11.7 & 0.22 & 0.21 & 0.20 & 0.19 & 0.18 &  &  \\
5Y into 2Y & 14.0 & 0.51 & 0.49 &  &  &  &  &  \\
10Y into 5Y & 10.0 & 0.22 & 0.21 & 0.20 & 0.19 & 0.18 &  &  \\
7Y into 5Y & 11.0 & 0.23 & 0.21 & 0.20 & 0.19 & 0.18 &  &  \\
10Y into 2Y & 12.2 & 0.51 & 0.49 &  &  &  &  &  \\
10Y into 7Y & 9.6 & 0.17 & 0.16 & 0.15 & 0.14 & 0.13 & 0.13 & 0.12 \\
2Y into 2Y & 14.8 & 0.52 & 0.48 &  &  &  &  &
\end{tabular}
\end{center}
\caption{Swaption volatilities and weights (data courtesy of BNP
Paribas, London).} \label{TabSwaptions}
\end{table}

\section{Numerical results}
\subsection{Price bounds}
We use a data set from Nov. 6 2000 and we plot in figure
(\ref{figsidney}) the upper and lower bounds obtained by
maximizing (resp. minimizing ) the volatility of a given swaption
provided that the Libor covariance matrix remains positive
semidefinite and that it matches the calibration data. We
calibrate by fitting all caplets up to 20 years plus the following
set of swaptions: 2Y into 5Y, 5Y into 5Y, 5Y into 2Y, 10Y into 5Y,
7Y into 5Y, 10Y into 2Y, 10Y into 7Y, 2Y into 2Y. This choice of
swaptions was motivated by liquidity (where all swaptions on
underlying and maturity in 2Y, 5Y, 7Y, 10Y are meant to be
liquid). For simplicity, all frequencies are annual. For each
$(Maturity,Underlying)$ pair in figure (\ref{figsidney}), we solve
the semidefinite program detailed in (\ref{StatCalib}):
\[
\BA{ll}
\mbox{find} & X \\
\mbox{s.t.} & \Tr(\Omega_{k}X)=\sigma_{k}^{2}T_{S_{k}}, \quad k=1,...,m \\
& X \succeq 0
\EA
\]
in the variable $X\in\symm^{20}$, where $\sigma_{k}$ and
$\Omega_{k}$ are computed as in (\ref{StatCalib}) using the data
in Table (\ref{TabCaplets}) and (\ref{TabSwaptions}). The programs
are solved using the SEDUMI code by \cite{Stur99}. We then compare
these upper and lower bounds (dotted lines) with the actual market
volatility (solid line). Quite surprisingly considering the
simplicity of the model (stationarity of the \textit{sliding}
Libor dynamics $L(t,\theta)$), figure (\ref{figsidney}) shows that
all swaptions seem to fit reasonably well in the bounds imposed by
the model, except for the 7Y and 10Y underlying. This is in line
with the findings of \cite{Long00}.

\subsection{Super-replication \& calibration stability}
Here, we compare the performance and stability of the various
calibration methodologies detailed here and in \cite{Rebo99}. For
simplicity, we neglect the different changes of measure between
forward measures and work with a multivariate lognormal model
where the underlying assets $x_{i,t}$ for $i=1,\ldots,5$ follow:
\BEQ
\label{simple-model} dx_{i,s}/x_{i,s}=\sigma_i
dW_s-\frac{1}{2}\sigma_i^T\sigma_i ds
\EEQ
where $W_s$ is a B.M. of dimension $5$. At time $0$, we set
$x_{i,0}=.1$ and the covariance matrix $\sigma^T\sigma$ is given
by:
\[
\left(\BA{ccccc}
0.0144 & 0.0133 & 0.0074 & 0.0029 & 0.0013 \\
0.0133 & 0.0225 & 0.0151 & 0.0063 & 0.0032 \\
0.0074 & 0.0151 & 0.0144 & 0.0065 & 0.0032 \\
0.0029 & 0.0063 & 0.0065 & 0.0081 & 0.0027 \\
0.0013 & 0.0032 & 0.0032 & 0.0027 & 0.0036 \\
\EA\right)
\]
We then compare the performance of a delta hedging strategy
implemented using various calibration techniques. In each case, we
hedge a short position in an ATM basket option with coefficients
$w_0=(0,0,.4,.1,.5)$, calibrating the model on all single asset
ATM calls and an ATM basket with weight $w_6=(1,1,0,0,0)$, the
weights $w_1,\ldots,w_5$ being then equal to the Euclidean basis.
All options have maturity one year and we rebalance the hedging
portfolio $33$ times over this period. At each time step, we
calibrate to these option prices computed using the model in
(\ref{simple-model}). To test the stability of the calibration
techniques, we add a uniformly distributed noise to the
calibration prices with amplitude equal to $\pm 10 \%$ of the
original price.

We use four different calibration techniques to get the covariance
matrices and compute the deltas. In the first one, we use the
exact model covariance above to compute baseline results. In the
second one, we use the simple stabilization technique detailed in
(\ref{tik-stabilization}) and solve:
\BEQ
\label{stab-calib}
\BA{ll}
\mbox{minimize} & \|X\|_2 \\
\mbox{s.t.} & \Tr(\Omega_{k}X) = \sigma_{k}^{2}T_{S_{k}},\quad k=1,...,6 \\
& X\succeq 0
\EA
\EEQ
in the variable $X$, where $\sigma_k$ and the matrices
$\Omega_{k}$ for $k=1,\ldots,6$ are computed from $w_1,\ldots,w_6$
as in (\ref{StatCalib}). In the third one, we use the super
replication technique detailed in~$\S \ref{sec:hedging-approx}$
and calibrate the covariance matrix by solving
\BEQ
\label{super-hedging-calib}
\BA{ll}
\mbox{maximize} &\Tr(\Omega_{0}X) \\
\mbox{s.t.} & \Tr(\Omega_{k}X) = \sigma_{k}^{2}T_{S_{k}},\quad k=1,...,6 \\
& X\succeq 0
\EA
\EEQ
in the variable $X$, where $\sigma_k$ and the matrices
$\Omega_{k}$ for $k=0,\ldots,6$ are computed from $w_0,\ldots,w_6$
as in (\ref{StatCalib}). Finally, our fourth calibrated covariance
matrix is calibrated using the two factor parametrized best fit
technique detailed in \cite{Rebo99}.

\begin{figure}[ht]
\begin{center}
\includegraphics[width=0.8 \textwidth]{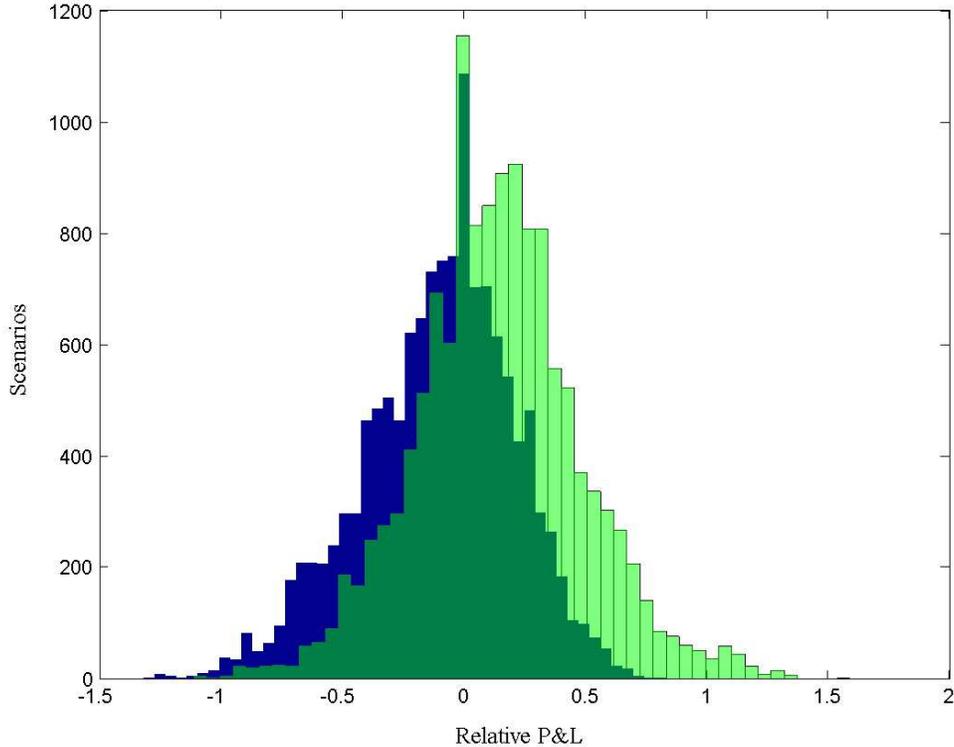}
\end{center}
\caption{P\&L distribution for the super-hedging (light) and the
parametric (dark) calibration techniques.
\label{fig-compare-super}}
\end{figure}

For each calibration technique, we record the ratio of the delta
hedging portfolio's P\&L to the initial option premium at every
time step. In figure (\ref{fig-compare-super}), we plot the P\&L
distributions for the super-hedging strategy
(\ref{super-hedging-calib}) and the \cite{Rebo99} calibration
technique. We notice that while discrete hedging error makes the
super-replication somewhat imperfect, the super-hedging
calibration technique has a much higher P\&L on average than the
best fit calibration in \cite{Rebo99}. Furthermore, the
super-hedging calibration produces a positive P\&L in 68\% of the
sample scenarios, while the parametrized calibration has a
positive P\&L in only 41\% of them.

Finally, in table (\ref{fig-compare-super}) we detail some summary
statistics on the relative P\&L of the four calibration techniques
detailed above and on the change in covariance matrix between time
steps. In this table, ``mean change'' is the average norm of the
change in the calibration matrix at each time step, while ``Mean
P\&L'' and ``StDev'' are the mean and standard deviation of the
ratio of the hedging portfolio's P\&L to the option premium.
Finally, the ``real covariance'' calibration technique uses the
actual model covariance matrix to compute the delta, the
``robust'' technique uses the solution to (\ref{stab-calib}), the
``super-hedging'' one uses the solution to
(\ref{super-hedging-calib}), while the ``parametrized'' uses the
parametric algorithm described in \cite{Rebo99}.

\begin{table}[ht]
\begin{center}
\begin{tabular}{ccccc}
\mbox{\bf Calibration method} & \mbox{Real covariance} &
\mbox{Robust} &
\mbox{Super-Hedging} & \mbox{Parameterized}\\
\hline \mbox{Mean P\&L} & -0.002 & 0.083 & 0.137 & -0.109 \\
\mbox{StDev P\&L} & 0.253 & 0.338 & 0.344 & 0.316 \\
\mbox{Mean change ($10^{-5}$)} & 0 & 3.11 & 3.36 & 4.77\\
\end{tabular}
\end{center}
\caption{Hedging P\&L and covariance stability statistics for
various calibration techniques.} \label{Tab-Stats-Hedging}
\end{table}

We notice that, as expected, the super hedging strategy improves
the average P\&L. However, while the robust calibration algorithm
does show a smaller average change in calibrated covariance
matrix, this does not translate into significantly smaller hedging
P\&L standard deviation. This is perhaps due to the fact that the
basket options considered here are not sensitive enough for these
changes in the covariance to have a significant impact on the
hedging performance.

\section{Conclusion}
The results above have showed how semidefinite programming based
calibration methods provide integrated calibration and
risk-management results with guaranteed numerical performance, the
dual program having a very natural interpretation in terms of
hedging instruments and sensitivity. Furthermore, these techniques
make possible the numerical computation of super-hedging
strategies (detailed in section \ref{sec:hedging-approx}), which
seem to perform well in our simulation examples.

\bibliographystyle{agsm}
\bibliography{jcf}
\end{document}

%% file: defs.tex
\newcommand{\BEAS}{\begin{eqnarray*}}
\newcommand{\EEAS}{\end{eqnarray*}}
\newcommand{\BEA}{\begin{eqnarray}}
\newcommand{\EEA}{\end{eqnarray}}
\newcommand{\BEQ}{\begin{equation}}
\newcommand{\EEQ}{\end{equation}}
\newcommand{\BIT}{\begin{itemize}}
\newcommand{\EIT}{\end{itemize}}
\newcommand{\BNUM}{\begin{enumerate}}
\newcommand{\ENUM}{\end{enumerate}}

\newcommand{\BA}{\begin{array}}
\newcommand{\EA}{\end{array}}

\newcommand{\ie}{{\it i.e.}}

\newcommand{\reals}{{\mbox{\bf R}}}

\newcommand{\symm}{{\mbox{\bf S}}}  

\newcommand{\Tr}{\mathop{\bf Tr}}

\newtheorem{theorem}{Theorem}

\newtheorem{remark}[theorem]{Remark}

\newcounter{exno}

{\begin{quote}\begin{small}\textbf{Proof.}}%
{\end{small}\end{quote}}

{\begin{quote}}{\end{quote}}

\makeatletter
\long\def\@makecaption#1#2{
   \vskip 9pt 
   \begin{small}
   \setbox\@tempboxa\hbox{{\bf #1:} #2}
   \ifdim \wd\@tempboxa > 5.5in
        \begin{center}
        \begin{minipage}[t]{5.5in}
        \addtolength{\baselineskip}{-0.95pt}
        {\bf #1:} #2 \par
        \addtolength{\baselineskip}{0.95pt}
        \end{minipage}
        \end{center}
   \else 
	\hbox to\hsize{\hfil\box\@tempboxa\hfil}  
   \fi
   \end{small}\par
}
\makeatother

\newcounter{oursection}

\newcounter{lecture}